\documentclass[a4paper,11pt]{article}
\usepackage{jheppub}
\usepackage{xspace}

\usepackage{todonotes}
\usepackage{lipsum}
\usepackage{comment}
\usepackage{mathtools}

\newcommand{\Libra}{\texttt{Libra}\xspace}
\newcommand{\LiteRed}{\texttt{LiteRed}\xspace}
\newcommand{\e}{\epsilon}

\newcommand{\eemumu}{\ensuremath{e^+e^-\to \mu^+\mu^-}\xspace}
\newcommand{\cA}{\mathcal{A}}
\newcommand{\cT}{\mathcal{T}}

\newcommand{\cH}{\mathcal{H}}
\newcommand{\cW}{\mathcal{W}}
\newcommand{\cV}{\mathcal{V}}
\newcommand{\cU}{\mathcal{U}}

\newcommand{\va}[1]{\left\langle#1\right\rangle}

\newcommand{\citegkl}{\cite{GKL2025}\xspace}

\title{Radiative correction to the charge asymmetry in $e^{+}e^{-}\to\mu^{+}\mu^{-}$ process.}
\author{Roman E. Gerasimov,}\emailAdd{r.e.gerasimov@inp.nsk.su}
\author{Petr A. Krachkov,}\emailAdd{p.a.krachkov@inp.nsk.su}
\author{and Roman N. Lee}\emailAdd{r.n.lee@inp.nsk.su}

\affiliation{Budker Institute of Nuclear Physics, Novosibirsk 630090, Russia}

\abstract{
	We calculate the next-to-next-to-leading order (NNLO) QED corrections to the $C$-odd part of the differential cross section of the $e^+e^-\to\mu^+\mu^-$ process. This part contributes to the angular and forward-backward asymmetry. Together with our earlier paper \cite{GKL2025}, this work completes the analytical calculation of $e^+e^-\to\mu^+\mu^-$ differential cross section at NNLO.
}
\begin{document}

\maketitle

\section{Introduction}
Muon pair production in electron-positron annihilation stands as one of the most fundamental processes in quantum electrodynamics (QED). Its simplicity in the Born approximation makes it the ``Hello, world!'' example in many quantum field theory textbooks. Nevertheless, this process also has a great importance for the experiments. It serves not only as a key process for luminosity determination at $e^+e^-$ colliders but also constitutes a significant background for searches of rare events and potential New Physics manifestations. The latter provides a strong motivation to make the theoretical predictions as precise as possible.

The experimental precision achieved at modern colliders necessitates theoretical predictions beyond the leading order. The next-to-leading order (NLO) QED corrections to this process have been known for decades~\cite{Berends1973, Berends1983,Jadach1984}. The pursuit of NNLO precision is now motivated by the requirements of ongoing and future experiments, as highlighted in recent reviews~\cite{Aliberti2024}. The calculation of the NNLO corrections involves the evaluation of two-loop virtual amplitudes and real radiative corrections. An important step towards the complete determination of NNLO differential cross section was done in our recent paper \citegkl. In that paper we have calculated the $C$-even part of the cross section exactly in the muon mass.

Although the muon mass $m_\mu$ can often be considered as small compared to the $\sqrt{s}$, we keep the $m_\mu$ dependence precise bearing in mind the applicability of the obtained results to the production of taus (with some reservations). The exact account of muon mass is also essential when this process serves as a background for the measurement of specific observables like the pion form factor contributing to the hadronic vacuum polarization correction to $(g-2)_\mu$. In contrast, the electron mass $m_e$ can be treated as a small parameter. However, one can not simply put it to zero  due to collinear divergences. These divergences must be regularized by keeping $m_e$ small but finite, which results to large logarithms of the form $\ln(s/m_e^2)$. The power corrections in $m_e$ can be neglected, of course.

\begin{figure}
	\includegraphics[width=\textwidth]{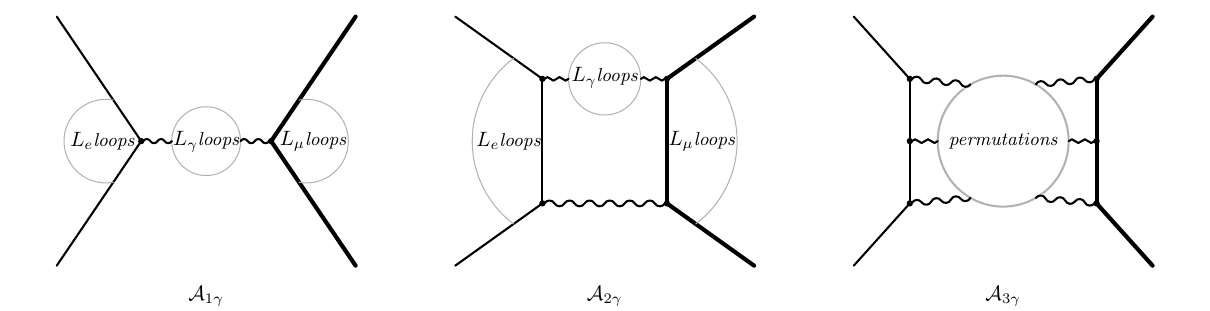}
	\caption{$1\gamma$-, $2\gamma$- and $3\gamma$-reducible diagrams that contribute to \eemumu process up to two-loops. First set contains one- and two-loop corrections to the electron and muon form factors and to the photon self-energy, $L_e+L_\mu+L_\gamma \leqslant2$. In the second diagram $L_e+L_\gamma+L_\mu\leqslant 1$.
	}
	\label{fig:NNLOdiagrams}
\end{figure}

The full NNLO amplitude can be decomposed into gauge-invariant sets of $n\gamma$-reducible diagrams, Fig. \ref{fig:NNLOdiagrams}. By definition, the $n\gamma$-reducible diagram is the one which contains $n$ intermediate photon lines. Similar to Ref. \citegkl, we will write $\mathcal{A}^{(L)}_{n\gamma}$ to denote the $L$-loop $n\gamma$-reducible diagrams.

The differential cross section of \eemumu process naturally splits into $C$-even and $C$-odd parts \citegkl,
\[d\sigma = d\sigma_{C\text{-even}} + d\sigma_{C\text{-odd}},\]
which are the functions of $c=\cos\theta$, $\theta$ is the scattering angle, of the corresponding parity (odd and even). At NNLO we have
\begin{align}
	d\sigma_{C\text{-even}}&=a^2\Big[|\cA_{1\gamma}^{(0)}|^2
	+2\Re \cA_{1\gamma}^{(1)} \cA_{1\gamma}^{(0)*}a
	+\Big(|\cA_{1\gamma}^{(1)}|^2+|\cA_{2\gamma}^{(1)}|^2\nonumber\\&
	\qquad\qquad\qquad+2\Re \cA_{1\gamma}^{(2)} \cA_{1\gamma}^{(0)*}
	+2\Re \cA_{3\gamma}^{(2)} \cA_{1\gamma}^{(0)*}\Big)a^2
	\Big]d\Phi,\label{eq:dsigma_even}\\
	d\sigma_{C\text{-odd}}&=2a^3\Re \left[
	\cA_{2\gamma}^{(1)} \cA_{1\gamma}^{(0)*}
	+\left(\cA_{2\gamma}^{(1)} \cA_{1\gamma}^{(1)*}
	+\cA_{2\gamma}^{(2)} \cA_{1\gamma}^{(0)*}\right)a
	\right]d\Phi\,,\label{eq:cs_odd}
\end{align}
where $a=\alpha/(4\pi)\approx (1/137.036)/(4\pi)$.
Here and below we use notation $Q^{(L)}$ to denote the coefficient of $L$-loop contribution to the quantity $Q$ in
\begin{equation}
	Q=a^{n_Q}\sum_{L=0}^{\infty} a^L Q^{(L)},
\end{equation}
where $a^{n_Q}$ is the parametric magnitude of tree contribution. In particular, we use this notation for $Q=\cA_{\bullet}, \cH_{\bullet}, Z_{\bullet}$.

The $C$-even part, symmetric under $\theta \to \pi - \theta$, was calculated in Ref.~\cite{GKL2025}. The present work focuses on the $C$-odd part, which is responsible for the asymmetric part of the cross section. Note that this asymmetry is absent at the QED Born level. Therefore, the leading contribution appears only at NLO, and the two-loop correction at NNLO becomes particularly important if compared to the correction to the $C$-even cross section. In Eq. \eqref{eq:cs_odd}, in addition to the amplitudes $\cA_{1\gamma}^{(0)},\ \cA_{1\gamma}^{(1)}$, and $\cA_{2\gamma}^{(1)}$, which already appeared in Ref. \citegkl, we have a new ingredient $\mathcal{A}^{(2)}_{2\gamma}$. The calculation of this amplitude is complicated by the appearance of electron mass logarithms in the second group of diagrams in Fig. \ref{fig:NNLOdiagrams} with $L_e=1$, $L_\mu=L_\gamma=0$. The corresponding master integrals have been computed in Ref.~\cite{Lee2025} using the Frobenius expansion in $m_e$.

Note that there are also corrections to the $C$-odd cross section which arise due to the account of electroweak and QCD effects. The former appears already at Born level when the interference between diagrams with intermediate photon and $Z$-boson exchange is taken into account. Being negligible at low energies, this contribution reaches the level of NNLO QED corrections already at the energy of 2 GeV. It is, of course, elementary to account this contribution. The QCD correction enters through the insertion of hadron vacuum polarization (HVP) in the one-loop box diagram and in the virtual photon propagator in the Born diagram with the radiation of soft photon. We will discuss the HVP contribution in Section~\ref{sec:had-2g-reducible}.

\section{Notations}

We use the same notations as in \citegkl. Let us summarize the key definitions here for self-consistence.

We consider the process
\begin{equation}
	e^-(p_1)+e^+(p_2) \longrightarrow \mu^-(q_1)+\mu^+(q_2)\,.
\end{equation}
and introduce the standard Mandelstam invariants
\begin{equation}
	s=(p_1+p_2)^2=(q_1+q_2)^2\,,\quad
	t=(p_1-q_1)^2=(p_2-q_2)^2\,,\quad
	u=(p_1-q_2)^2=(p_2-q_1)^2\,.
\end{equation}
The momenta and invariants satisfy the usual constraints
\begin{equation}
	p_1+p_2=q_1+q_2\,,\quad
	p_1^2=p_2^2=m^2\,,\quad
	q_1^2=q_2^2=M^2\,,\quad
	s+t+u =2m^2+2M^2\,,
\end{equation}
where $m=m_e$ and $M=m_\mu$ are the electron and muon masses, respectively.
Throughout the paper we also use the notations
\begin{equation}
\beta = \sqrt{1-\tfrac{4M^2}{s}}\,,\quad c=\cos\theta\,,
\end{equation}
which are the muon velocity and cosine of the angle between electron and muon momenta in c.m. frame. The Mandelstam variables in terms of $\beta$ and $c$ are
\begin{equation}
s=\frac{4M^2}{1-\beta^2}\,,\quad t=M^2\left(1-2\frac{1-\beta c}{1-\beta^2}\right)\,,\quad u=M^2\left(1-2\frac{1+\beta c}{1-\beta^2}\right)\,,
\end{equation}
where we neglect the electron mass.

\begin{figure}
	\includegraphics[width=\textwidth]{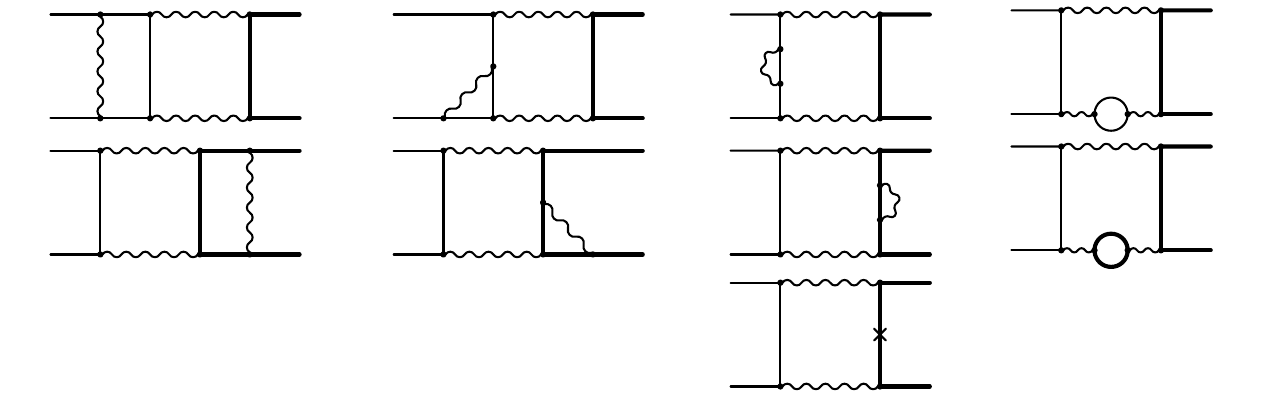}
	\caption{Representative diagrams contributing to the amplitude $\mathcal{A}^{(2)}_{2\gamma}$. The diagram in the third line corresponds to the contribution $\cA^{(1)}_{2\gamma,\text{mct}}$ due to the mass counterterm. The cross denotes the vertex $-iM$.}
	\label{fig:diagrams}
\end{figure}
The diagrams for QED contributions to the amplitude  $\mathcal{A}^{(2)}_{2\gamma}$ are shown in Fig.~\ref{fig:diagrams}. The diagrams in the first line can be expressed via master integrals calculated in Ref.~\cite{Lee2025}\footnote{Note that the definition of $t$ in~\cite{Lee2025} differs by the sign of $\cos{\theta}$.}. The results of Ref.~\cite{Lee2025} have the form of Frobenius series in the electron mass and therefore assume that $m\ll M$. Thus the diagrams in the second line of Fig.~\ref{fig:diagrams} can not be obtained from those by the replacement $m\leftrightarrow M$ and should be calculated separately. Fortunately, these diagrams do not contain collinear divergences and we  can set $m=0$ to essentially simplify the calculation. We present some details of the derivation of the corresponding master integrals in the Appendix~\ref{sec:basis-Lmu-1}.

\section{Hard amplitudes}
We use dimensional regularization, $d=4-2\epsilon$, to regularize both UV and IR divergences.
In order to eliminate the UV divergences, we perform the on-shell renormalization procedure. Denoting by $\cA^{(2)}_{2\gamma,\text{bare}}$ the sum of the diagrams in the first two lines of Fig. \ref{fig:diagrams}, we have
\begin{equation}
\cA^{(2)}_{2\gamma} = \cA^{(2)}_{2\gamma,\text{bare}} +\left[2Z^{(1)}_A(m)+2Z^{(1)}_A(M)-Z^{(1)}_{\psi}(m)-Z^{(1)}_{\psi}(M)\right]\cA^{(1)}_{2\gamma}
+Z^{(1)}_m(M)\cA^{(1)}_{2\gamma,\text{mct}},
\end{equation}
where
\begin{align}
	Z^{(1)}_A=-\frac{4}{3\epsilon}\left(\frac{4\pi}{m^2}\right)^{\epsilon}\Gamma(1+\epsilon)\,,\quad
	Z^{(1)}_m=	Z^{(1)}_\psi=-\frac{3-2\epsilon}{\epsilon(1-2\epsilon)}\left(\frac{4\pi}{m^2}\right)^{\epsilon}\Gamma(1+\epsilon)\,,
\end{align}
are the one-loop on-shell renormalization constants and $\cA^{(1)}_{2\gamma,\text{mct}}$ denotes the contribution due to mass counterterm depicted in the last line of Fig. \ref{fig:diagrams}. Note that the corresponding diagram with cross on the electron line can be neglected within our precision as it is proportional to $m$.

After the renormalization the amplitude still contains the soft IR divergences which can be factorized as \cite{Yennie1961}
\begin{equation}
	\cA = e^{a\cV} \cH\,,\label{eq:AviaH}
\end{equation}
where \citegkl
\begin{align}\label{eq:Vif}
	\cV&=\cV_{II}+\cV_{FF}+\cV_{IF}
	=V(p_1,p_2)+V(-q_1,-q_2)+2[V(p_1,-q_1)-V(p_1,-q_2)].
\end{align}
with
\begin{equation}
	V\left(p_{i},p_{j}\right)=
	-8\pi^2
	\int\frac{\underline{d}^{d}k}{i\left(2\pi\right)^{d}}\frac{1}{k^{2}+i0}\left(\frac{2p_{i}-k}{k^{2}-2kp_{i}+i0}+\frac{2p_{j}+k}{k^{2}+2kp_{j}+i0}\right)^{2}.\label{eq:SVfunction}
\end{equation}
The explicit form of the function $V\left(p_{i},p_{j}\right)$ can be found in Refs. \cite{Fadin:2023phc,GKL2025}. The quantity $\cH$ in~\eqref{eq:AviaH}, dubbed the \emph{hard amplitude}, is finite at $\epsilon=0$.
For two-loop contribution, we have
\begin{align}
	\cH_{2\gamma}^{(2)} &= \cA_{2\gamma}^{(2)}-\cV_{IF}\cA_{1\gamma}^{(1)}-\left(\cV_{II}+\cV_{FF}\right)\cA_{2\gamma}^{(1)}+\cV_{IF}\left(\cV_{II}+\cV_{FF}\right)\cA_{1\gamma}^{(0)}\,.
\end{align}

As shown in Ref. \citegkl, the general form of the hard amplitude can be written as
\begin{equation}
	\cH= H_{1} \cT_1 +H_{3} \cT_3 + H_{4}  \cT_4\,,\label{eq:Hform}
\end{equation}
where the coefficients $H_{i}$ ($i=1,3,4$) play the role of invariant amplitudes and
\begin{equation}
	\cT_1 = \overline{U}V \,\bar v (\hat q_1-\hat q_2) u,\quad\
	\cT_3 = \overline{U}(\hat p_1-\hat p_2)V \,\bar v (\hat q_1-\hat q_2) u,\quad\
	\cT_4 = \overline{U}\gamma_{\mu}V \,\bar v\gamma^{\mu}u.
\end{equation}

Using the results for the required master integrals from Ref. \cite{Lee2025} and from Appendix \ref{sec:basis-Lmu-1}, we obtain the expressions\footnote{As we mentioned earlier, we neglect the terms suppressed by the power of $m$.} of the invariant amplitudes $H_{i,2\gamma}^{(2)}$ ($i=1,3,4$) in terms of the Goncharov's polylogarithms $G(\boldsymbol{w}|\beta)$. The weights $w_k$ belong to the alphabet
\begin{equation}
	\{0,\pm1,\pm c,\pm 1/c, \pm e^{i\theta}, \pm e^{-i\theta}\}.\label{eq:alphabet}
\end{equation}

It is interesting to note that the weights $\pm c$ appear due to the diagrams on the second line of Fig. \ref{fig:diagrams}. These weights result in the branching of individual polylogarithms on the lines $\beta=\pm c$ inside the physical region. However, we have checked that these branchings cancel in the resulting expression, which turns out to be analytic in the vicinity of these lines\footnote{Of course, the analyticity in the vicinity of $\beta=\pm c$ holds only for the physical sheet, therefore, it is impossible to fully avoid the weight $\pm c$ in the individual polylogarithms.}\!.

The found expressions for $H_{i,2\gamma}^{(2)}$ ($i=1,3,4$) are available in computer-readable form in ancillary files, see Section \ref{sec:ancillary} for description.

\subsection{Asymptotics}

Using our exact results, we calculate the threshold asymptotics and high-energy asymptotics.
The calculation of the threshold asymptotics is straightforward as it is reduced to the expansion of Goncharov's polylogarithms at small argument.
The calculation of the high-energy asymptotics is more involved. First, we pass from polylogarithms of argument $\beta$ to those of argument $1-\beta$. When we do so, the coefficients in front of $G(\ldots|1-\beta)$ are expressed in terms of $G(\boldsymbol{w}|1)$ where weights belong to the same alphabet \eqref{eq:alphabet}. Then the functions $G(\ldots|1-\beta)$ can be easily expanded in $1-\beta$ or, equivalently, in $1/s=(1-\beta^2)/4$. Expanding up to $1/s^4$, we collect the terms by the inverse powers of $s$ and by linearly independent rational functions of $c$. Specifically, for our purpose it is sufficient to take the functions $(1\pm c)^{-n}$ with $n\in \{0,1,2,3\}$. These functions are multiplied by the linear combinations with integer coefficients of $G(\boldsymbol{w}|1)$ with $w_k$ from \eqref{eq:alphabet}. In order to express these combinations in terms of the classical polylogarithms with argument depending on $c$, it is possible to use the symbol map. However, we find it simpler to use the following practical approach. We determine a conjectural basis of polylogarithmic functions of $c$ which may be involved. We use the basis of functions which appeared in the high-energy asymptotics of the $C$-even part of the cross section obtained in our paper \citegkl. It is constructed of (the products of) the functions
\begin{equation}
	\ln\tfrac{1\pm c}2,\, \ln 2,\,\mathrm{Li}_2\left(\tfrac{1+c}{2}\right),\, \pi^2,
	\mathrm{Li}_3\left(\tfrac{1\pm c}{2}\right),\,\zeta_3,\,
	\mathrm{Li}_4\left(\tfrac{1\pm c}{2}\right),\,\mathrm{Li}_4\left(\tfrac{1-c}{1+c}\right)
\,.
\end{equation}
Then we choose $\theta$ so as to make $c=\cos{\theta}$ and $e^{i\theta}$ sufficiently transcendental (e.g., we can take $\theta=1$) and use PSLQ algorithm to recognize the required linear combinations of $G(\ldots|1)$ in terms of linear combinations of the described basis of functions with rational coefficients. For more safety, we numerically check the obtained \emph{analytic} expressions in several other points with high precision.

\subsection{Hadronic vacuum polarization contribution}
\label{sec:had-2g-reducible}
\begin{figure}
	\centering
	\includegraphics[width=0.3\linewidth]{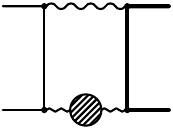}
	\caption{Contribution of hadronic vacuum polarization to $\cA_{2\gamma}$ amplitude. The blob denotes the insertion of the hadronic polarization operator.}
	\label{fig:HVP}
\end{figure}

The hadronic vacuum polarization enters the NNLO corrections to the amplitude via the diagram shown in Fig. \ref{fig:HVP}. We will denote the corresponding contribution to the amplitude as $\cA_{2\gamma, \text{(had)}}^{(2)}$.

In order to evaluate this contribution to $C$-odd radiative corrections we use the subtracted dispersion relation for the renormalized hadronic polarization operator:
\begin{equation}
	\label{eq:Pi-had-dispersion}
	\Pi^{\text{(had)}}(q^2) =
	\frac{q^2}{\pi}\int_{0}^{+\infty}\frac{d\Lambda^2}{\Lambda^2}\,\frac{\Im{\Pi^{\text{(had)}}(\Lambda^2)}}{\Lambda^2-q^2-i0}\,,
\end{equation}
This polarization operator corresponds to the on-shell renormalization of the photon propagator.

\begin{figure}
	\centering
	\includegraphics[width=.4\textwidth]{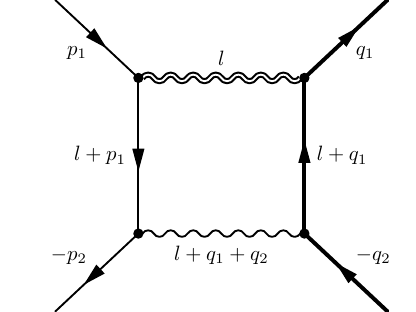}
	\caption{The Feynman diagrams that contribute to the amplitude \(\cA_{\Lambda^{2}}\); the double wavy line with momentum $l$ represents the massive particle with propagator $\frac{-i g_{\mu\nu}}{l^2-\Lambda^2 + i0}$.}
	\label{fig:box-lambda}
\end{figure}

The hadronic polarization operator~\eqref{eq:Pi-had-dispersion}, when sandwiched between two photon propagators, yields
\begin{equation}
	\label{eq:Pi-insertion}
	\begin{split}
		\frac{(-i g_{\mu\nu})}{l^2} \to \frac{(-i g_{\mu\rho})}{l^2}\times
		(i\Pi^{\text{(had)}}(l^2))\left(l^2g^{\rho\sigma} - l^{\rho}l^{\sigma}\right)\times
		\frac{(-i g_{\sigma\nu})}{l^2} \to \\
		\to -\frac{1}{\pi} \int_{s_{0}}^{+\infty}\frac{d{\Lambda^2}}{\Lambda^2}\,\Im{\Pi^{\text{(had)}}(\Lambda^2)}\ \frac{(-i g_{\mu\nu})}{l^2-\Lambda^2 + i0}
		~,
	\end{split}
\end{equation}
where we omit the terms proportional $l_{\mu}l_{\nu}$ taking into account leptonic ``currents'' conservation. The insertion of~\eqref{eq:Pi-insertion} leads to the following representation of the corresponding $2\gamma$-reducible amplitude
\begin{equation}
	\label{eq:2gamma-had-repr}
	\cA_{2\gamma, \text{(had)}}^{(2)} = -\frac{1}{\pi} \int_{0}^{+\infty} \frac{d \Lambda^2}{\Lambda^2}\,
	\frac{\Im{\Pi^{\text{(had)}}(\Lambda^2)}}{a}
	\,\cA_{\Lambda^{2}}^{(1)}
	~,
\end{equation}
where the amplitude \(\cA_{\Lambda^{2}}^{(1)}\) is associated with the four box diagrams.
One of the diagrams is presented in Fig.~\ref{fig:box-lambda}, and others correspond to permutations $p_1 \leftrightarrow p_2$ and/or $q_1 \leftrightarrow q_2$.

The hard amplitude corresponding to $\cA_{2\gamma,\text{(had)}}^{(2)}$ reads
\begin{equation}
	\cH_{2\gamma,\text{(had)}}^{(2)} =
	\cA_{2\gamma,\text{(had)}}^{(2)} - \cV_{IF}\cA_{1\gamma,\text{(had)}}^{(1)}\,,
\end{equation}
where
\begin{equation}
	\label{eq:AandH11had}
	\cA_{1\gamma,\text{(had)}}^{(1)} = \cH_{1\gamma,\text{(had)}}^{(1)} = \frac{\Pi^{\text{(had)}}}{a}\, \cA_{1\gamma}^{(0)}\,.
\end{equation}
Using Eqs. \eqref{eq:Pi-had-dispersion} and \eqref{eq:2gamma-had-repr}, we obtain 
\begin{align}
	\cH_{2\gamma,\text{(had)}}^{(2)} &=
	-\frac{1}{\pi} \int_{0}^{+\infty} \frac{d \Lambda^2}{\Lambda^2}\,\frac{\Im{\Pi^{\text{(had)}}(\Lambda^2)}}{a}\cH_{\Lambda^2}^{(1)}\label{eq:HhadIntegral}\\
	\cH_{\Lambda^2}^{(1)}&=
	\cA_{\Lambda^{2}}^{(1)} - \cV_{IF} \cA_{\Lambda^2}^{(0)}
	=\cA_{\Lambda^{2}}^{(1)} - \cV_{IF}\tfrac{s}{s-\Lambda^2+i0}\cA_{1\gamma}^{(0)}\,.\label{eq:Hhad}
\end{align}

The imaginary part of the polarization operator which enters the integrand in Eq.~\eqref{eq:2gamma-had-repr} can be extracted with required accuracy from the ratio $R(s)$ of total cross sections for hadrons and muon pairs production in electron-positron annihilation:
\begin{equation}
	\label{eq:R-s}
	\frac{\Im\Pi^{\text{(had)}}(s)}{a} =
	- \frac{4}{3} R(s) =
	- \frac{4}{3}
	\frac{\sigma_{e^{+ } e^{- } \to \text{hadrons}}}
	{\sigma_{e^{+ } e^{- } \to \mu^{+ }\mu^{- }}}
	~.
\end{equation}

In order to evaluate $\cA_{\Lambda^{2}}^{(1)}$, we consider the family
\begin{equation}
	\label{eq:box-lambda-family}
	\begin{split}
		j_{\text{box}}(n_{1},n_{2},n_{3},n_{4}) =
		\int \frac{d^{d}l}{i \pi^{d/2}}
		\prod_{k=1}^{4}\frac{1}{D_{\text{box},k}^{n_{k}}}
		~,
	\end{split}
\end{equation}
with the following sets of denominators
\begin{equation}
	\label{eq:box-denominators}
	\begin{split}
		&D_{\text{box},1} = -(l + p_1)^2-i0~,\quad
		D_{\text{box},2} = M^2-(l + q_1)^2-i0~,\\
		&D_{\text{box},3} = \Lambda^2-l^2-i0~,\quad
		D_{\text{box},4} = -(l + q_1 + q_2)^2-i0~.
	\end{split}
\end{equation}
The IBP reduction yields 8 master integrals, which can be taken from the literature. In particular, we use the intermediate results of Ref. \cite{Ignatov:2022iou}.
The one-loop amplitude $\cA_{\Lambda^{2}}^{(1)}$ can be expressed in terms of these 8 integrals together with the integrals obtained from them by $t\leftrightarrow u$.
Then, using Eq. \eqref{eq:Hhad}, we find the integration kernels $H^{(1)}_{i,\Lambda^2}$ ($i=1,3,4$) in tensor decomposition of $\cH^{(1)}_{\Lambda^2}$,
\begin{equation}
\cH^{(1)}_{\Lambda^2}=H^{(1)}_{1,\Lambda^2}\cT_1+H^{(1)}_{3,\Lambda^2}\cT_3+H^{(1)}_{4,\Lambda^2}\cT_4\,.
\label{eq:HLambda2}
\end{equation}

The found expressions for $H^{(1)}_{i,\Lambda^2}$ are available in computer-readable form in ancillary files, see Section \ref{sec:ancillary} for description.

\section{Relative corrections to $d\sigma_{C\text{-odd}}$ and asymmetries}

The differential cross section which includes the radiation of soft photons is expressed via hard amplitude as \citegkl
\begin{equation}
\begin{split}
	d\sigma_{\text{incl}}(\omega_0) &= e^{a\cU}\left|\cH\right|^2d\Phi.\\
	\cU=\cU(\omega_0)&=\cW(\omega_0)+2\Re \cV,
	\label{eq:csviaH}
\end{split}
\end{equation}
where
\begin{equation}
	d\Phi =\frac{8(2\pi)^2}{s^2}\, \delta(p_1+p_2-q_1-q_2)\, d^3{q}_1\,d^3{q}_2
\end{equation}
is the phase space in $d=4$ dimensions and the quantity $\cW(\omega_0)$ determines the dimensionally regularized probability of irradiation of soft photon with energy less than $\omega_0$. In what follows, to reduce clutter, we drop the `incl' subscript on cross sections and indicate inclusive cross sections only through their dependence on $\omega_0$.

The quantity $\cW(\omega_0)$ reads
\begin{align}\label{eq:Wif}
	\cW(\omega_0)&=\cW_{II}(\omega_0)+\cW_{FF}(\omega_0)+\cW_{IF}(\omega_0)\nonumber\\
	&=W(p_1,p_2|\omega_0)+W(-q_1,-q_2|\omega_0)+2[W(p_1,-q_1|\omega_0)-W(p_1,-q_2|\omega_0)].
\end{align}
The function $W(p_i,p_j|\omega_0)$ is defined as
\begin{equation}
	W(p_i,p_j|\omega_0)=-16\pi^2\left(\frac{e^{\gamma_E}}{4\pi}\right)^\e\intop_{\omega<\omega_0}\frac{d^{d-1}k}{(2\pi)^{d-1}2\omega}\left(\frac{p_i}{k\cdot p_i}-\frac{p_j}{k\cdot p_j}\right)^2
	~.\label{eq:Wint}
\end{equation}
Its expansion up to $\e^0$ can be found in Ref. \citegkl. It is important that $\cU$, as well as specific contributions to $\cU$, like $\cU_{IF}=\cW_{IF}(\omega_0)+2\Re \cV_{IF}$, are all finite at $\e\to 0$.

Although our results allow to consider also the polarization effect in the differential cross section, we restrict ourselves below to the unpolarized case. Passing to unpolarized  cross section amounts to the replacement $\left|\cH\right|^2\to \overline{\Sigma}\left|\cH\right|^2$ in Eq. \eqref{eq:csviaH}, where $\overline{\Sigma}$ denotes averaging/summing over polarizations of initial/final particles. In what follows we will imply $\overline{\Sigma}$ in front of any quadratic combinations of hard amplitudes, but will not write it explicitly to lighten the formulas.

We define relative corrections $\delta^{(L)}_{C\text{-odd}}\propto a^L$ to the cross section $d\sigma_{C\text{-odd}}$ via
\begin{equation}
	\frac{d\sigma_{C\text{-odd}}(\omega_0)}{d\Omega} =\frac{d\sigma_0}{d\Omega}\left[\delta^{(1)}_{C\text{-odd}} + \delta^{(2)}_{C\text{-odd}} + \delta^{(2)}_{C\text{-odd},\text{(had)}}+O(a^3)\right]\,,
\end{equation}
where
\begin{equation}
	\frac{d\sigma_0}{d\Omega} = \frac{\alpha^2\beta}{4s}\left[2-\beta^2(1-c^2)\right]
\end{equation}
is the differential Born cross section.

Note that the contribution of hadronic vacuum polarization to the $C$-odd cross section appears for the first time at two loops. In terms of hard amplitudes, these  relative corrections read

\begin{equation}
	\label{eq:delta-c-odd-1}
	\delta_{\text{C-odd}}^{(1)} = a\Bigg[
	\tfrac{2\Re\left[\cH_{2\gamma}^{(1)}\cH_{1\gamma}^{(0)*}\right]}{\left|\cH_{1\gamma}^{(0)}\right|^2}
	+ \cU_{IF}
	\Bigg]\,,
\end{equation}
\begin{multline}
	\label{eq:delta-c-odd-2}
	\delta_{\text{C-odd}}^{(2)} = a^2 \Bigg[
	\tfrac{2\Re\left[\cH_{2\gamma}^{(1)}\cH_{1\gamma}^{(1)*}\right]}{\left|\cH_{1\gamma}^{(0)}\right|^2} + \tfrac{2\Re\left[\cH_{2\gamma}^{(2)}\cH_{1\gamma}^{(0)*}\right]}{\left|\cH_{1\gamma}^{(0)}\right|^2}
\\
+ \cU_{IF} \tfrac{2\Re\left[\cH_{1\gamma}^{(1)}\cH_{1\gamma}^{(0)*}\right]}{\left|\cH_{1\gamma}^{(0)}\right|^2}
	+ (\cU_{II} + \cU_{FF}) \left(
		\tfrac{2\Re\left[\cH_{2\gamma}^{(1)}\cH_{1\gamma}^{(0)*}\right]}{\left|\cH_{1\gamma}^{(0)}\right|^2}
		+\cU_{IF}
		\right)
	\Bigg]\,,
\end{multline}
and
\begin{equation}
	\label{eq:delta-c-odd-2-had}
	\delta_{\text{C-odd},\text{(had)}}^{(2)} = a^2 \Bigg[
		\tfrac{2\Re\left[\cH_{2\gamma}^{(1)}\cH_{1\gamma,\text{(had)}}^{(1)*}\right]}{\left|\cH_{1\gamma}^{(0)}\right|^2}
		 + \tfrac{2\Re\left[\cH_{2\gamma,\text{(had)}}^{(2)}\cH_{1\gamma}^{(0)*}\right]}{\left|\cH_{1\gamma}^{(0)}\right|^2}
		 + \cU_{IF} \tfrac{2\Re\left[\cH_{1\gamma,\text{(had)}}^{(1)}\cH_{1\gamma}^{(0)*}\right]}{\left|\cH_{1\gamma}^{(0)}\right|^2}
	\Bigg]\,.
\end{equation}

\subsection{Asymmetry}
The differential angular asymmetry is defined via
\begin{equation}
	\mathrm{A}(c)=\frac{d\sigma_{C\text{-odd}}(\omega_0)}{d\sigma_{C\text{-even}}(\omega_0)}
	\label{eq:As},
\end{equation}
so that $\mathrm{A}(c)=-\mathrm{A}(-c)$.
In the leading approximation the denominator in Eq. \eqref{eq:As} can be replaced by $d\sigma_0$ and the asymmetry coincides with the $C$-odd relative correction $\delta_{\text{C-odd}}^{(1)}$, Eq. \eqref{eq:delta-c-odd-1},
\begin{equation}
	\mathrm{A}^{(1)}(c) = \delta_{\text{C-odd}}^{(1)}= a\Bigg[
	\tfrac{2\Re\left[\cH_{2\gamma}^{(1)}\cH_{1\gamma}^{(0)*}\right]}{\left|\cH_{1\gamma}^{(0)}\right|^2}+ \cU_{IF}
	\Bigg]\,.\label{eq:A1}
\end{equation}
In particular, $\mathrm{A}^{(1)}(c)$ depends on $\omega_0$ via $\cU_{IF}$ in Eq. \eqref{eq:delta-c-odd-1}. In order to find the $O(a^2)$ correction to the asymmetry, we need to take into account the expansion of the denominator up to $a^1$.
The $C$-even cross section with the required accuracy reads
\begin{equation}
	\frac{d\sigma_{C\text{-even}}(\omega_0)}{d\Omega} =\frac{d\sigma_0}{d\Omega}\left[1+\delta^{(1)}_{C\text{-even}} + \delta^{(1)}_{C\text{-even},\text{(had)}} +O(a^2)\right]\,.
\end{equation}
where
\begin{equation}
	\label{eq:delta-c-even}
	\delta_{\text{C-even}}^{(1)} = a \left[ \tfrac{2\Re\left[\cH_{1\gamma}^{(1)}\cH_{1\gamma}^{(0)*}\right]}{\left|\cH_{1\gamma}^{(0)}\right|^2}
	+\cU_{II}+\cU_{FF}
	\right],\qquad\quad
	\delta_{\text{C-even},\text{(had)}}^{(1)} = a
	\tfrac{2\Re\left[\cH_{1\gamma,\text{(had)}}^{(1)}\cH_{1\gamma}^{(0)*}\right]}{\left|\cH_{1\gamma}^{(0)}\right|^2}\,.
\end{equation}
Then we obtain
\begin{equation}
	\mathrm{A}^{(2)}(c) + \mathrm{A}_{\text{(had)}}^{(2)}(c) = \left[\delta_{\text{C-odd}}^{(2)}
	-\delta_{\text{C-odd}}^{(1)}\delta_{\text{C-even}}^{(1)}\right]
	 + \left[\delta_{\text{C-odd},\text{(had)}}^{(2)}-\delta_{\text{C-odd}}^{(1)}\delta_{\text{C-even},\text{(had)}}^{(1)}\right]\,.
\end{equation}
Substituting Eqs. \eqref{eq:delta-c-odd-1}, \eqref{eq:delta-c-odd-2}, \eqref{eq:delta-c-odd-2-had}, and \eqref{eq:delta-c-even}, we obtain
\begin{align}
	\mathrm{A}^{(2)}(c)&=a^2\Bigg[\tfrac{2\Re\left[\mathcal{H}^{(2)}_{2\gamma}\mathcal{H}^{(0)*}_{1\gamma}\right]}{\left|\cH_{1\gamma}^{(0)}\right|^2}+\tfrac{2\Re\left[\mathcal{H}^{(1)}_{2\gamma}\mathcal{H}^{(1)*}_{1\gamma}\right]}{\left|\cH_{1\gamma}^{(0)}\right|^2}-\tfrac{2\Re\left[\mathcal{H}^{(1)}_{2\gamma}\mathcal{H}^{(0)*}_{1\gamma}\right]}{\left|\cH_{1\gamma}^{(0)}\right|^2} \tfrac{2\Re\left[\mathcal{H}^{(1)}_{1\gamma}\mathcal{H}^{(0)*}_{1\gamma}\right]}{\left|\cH_{1\gamma}^{(0)}\right|^2}\Bigg]\,,\nonumber
	\\
	\mathrm{A}_{\text{(had)}}^{(2)}(c)&=\mathrm{A}^{(2)}\bigg|_{\cH_{1\gamma}^{(1)}\to \cH_{1\gamma,\text{(had)}}^{(1)},\ \cH_{2\gamma}^{(2)}\to\cH_{2\gamma,\text{(had)}}^{(2)}}
	\label{eq:A2}
\end{align}
Remarkably, the terms depending on $\cU$ canceled and $\mathrm{A}^{(2)}$ and $\mathrm{A}_{\text{(had)}}^{(2)}$ do not depend on $\omega_0$.

Another conventional experimental quantity is the \textit{forward-backward asymmetry} $\mathrm{A}_{\text{FB}}$ defined as
\begin{equation}
	\mathrm{A}_{\text{FB}}=\frac{\int\limits_{c>0} d\sigma-\int\limits_{c<0}d\sigma}{\int\limits_{c>0}d\sigma+\int\limits_{c<0}d\sigma}
	=\frac{\int\limits_{c>0} d\sigma_{C\text{-odd}}(\omega_{0})}{\int\limits_{c>0} d\sigma_{C\text{-even}}(\omega_{0})}
	\,.\label{eq:AFB}
\end{equation}
At first glance, it may seem that $\mathrm{A}_{\text{FB}}$ at one and two loops can be given by wrapping each quadratic in $\cH$ combination in \eqref{eq:A1} and \eqref{eq:A2} with the integration $\intop_0^1dc[\ldots]$. Introducing weighted angle averaging $\va{\ldots} = \int_{c>0} \frac{2d\sigma_{0}}{\sigma_0}$ we obtain at one loop
\begin{equation}
		A_{\text{FB}}^{(1)} = \va{A^{(1)}(c)}
		= A_{\text{FB}}^{(1,\text{hard})} +A_{\text{FB}}^{(1,\text{soft})}
		=a\va{\tfrac{2\Re\left[\cH_{2\gamma}^{(1)}\cH_{1\gamma}^{(0)*}\right]}{\left|\cH_{1\gamma}^{(0)}\right|^2}}
		+ a\va{\cU_{IF}}.
\end{equation}
In particular, we have
\begin{equation}
	A_{\text{FB}}^{(1,\text{soft})} = \frac{3a}{2(3-\beta^2)}\int_{0}^{1}(2-\beta^2(1-c^2))\,\mathcal{U}_{IF}\,dc
\end{equation}
For two loops the same prescription gives
\begin{equation}
	\mathrm{A}_{\text{FB}}^{(2,\text{hard})}=a^2\Bigg[
	\va{\tfrac{2\Re\left[\mathcal{H}^{(2)}_{2\gamma}\mathcal{H}^{(0)*}_{1\gamma}\right]}{\left|\cH_{1\gamma}^{(0)}\right|^2}}
	+\va{\tfrac{2\Re\left[\mathcal{H}^{(1)}_{2\gamma}\mathcal{H}^{(1)*}_{1\gamma}\right]}{\left|\cH_{1\gamma}^{(0)}\right|^2}}
	-\va{\tfrac{2\Re\left[\mathcal{H}^{(1)}_{2\gamma}\mathcal{H}^{(0)*}_{1\gamma}\right]}{\left|\cH_{1\gamma}^{(0)}\right|^2}} \va{\tfrac{2\Re\left[\mathcal{H}^{(1)}_{1\gamma}\mathcal{H}^{(0)*}_{1\gamma}\right]}{\left|\cH_{1\gamma}^{(0)}\right|^2}}\Bigg]
\end{equation}
However, this is not the only contribution to $\mathrm{A}_{\text{FB}}^{(2)}$ because the functions $\cU_{IF}$ depend on $c$ and enter the integrals over $c$ in the numerator and denominator of Eq. \eqref{eq:AFB} with different integration weights and, thus, their contribution does not vanish. The correct expression for $\mathrm{A}_{\text{FB}}^{(2)}$ is
\begin{equation}
	\mathrm{A}_{\text{FB}}^{(2)}=\mathrm{A}_{\text{FB}}^{(2,\text{hard})}+\mathrm{A}_{\text{FB}}^{(2,\text{soft})},
\end{equation}
where
\begin{align}
	\mathrm{A}_{\text{FB}}^{(2,\text{soft})} &= a^2 \left[\va{\cU_{IF} \tfrac{2\Re\left[\mathcal{H}_{1\gamma}^{(1)}\mathcal{H}_{1\gamma}^{(0)*}\right]}{\left|\mathcal{H}_{1\gamma}^{(0)}\right|^2}} - \va{\cU_{IF} }\va{\tfrac{2\Re\left[\mathcal{H}_{1\gamma}^{(1)}\mathcal{H}_{1\gamma}^{(0)*}\right]}{\left|\mathcal{H}_{1\gamma}^{(0)}\right|^2}}\right]\nonumber\\
	&=\frac{3 a^2 \beta (1-\beta^2)}{(3-\beta^2)^2}\
	\ln\left(\frac{1+\beta}{1-\beta}\right)\
	\int_{0}^{1}(3c^2-1)\,\cU_{IF}\, dc
\end{align}

The contributions to forward-backward asymmetry related to the hadronic vacuum polarization are obtained by the same replacement $\cH_{1\gamma}^{(1)}\to \cH_{1\gamma,\text{(had)}}^{(1)},\ \cH_{2\gamma}^{(2)}\to\cH_{2\gamma,\text{(had)}}^{(2)}$ as in Eq. \eqref{eq:A2} followed by deletion of the terms which do not depend on $\cH_{n\gamma,\text{(had)}}^{(L)}$. In particular, we have
\begin{equation}
\begin{split}
	&A_{\text{FB},\text{(had)}}^{(1)} = 0~,\quad
         A_{\text{FB},\text{(had)}}^{(2,\text{soft})} = 0~,\\
        &A_{\text{FB},\text{(had)}}^{(2,\text{hard})} = A_{\text{FB}}^{(2,\text{hard})}\bigg|_{\cH_{1\gamma}^{(1)}\to \cH_{1\gamma,\text{(had)}}^{(1)},\ \cH_{2\gamma}^{(2)}\to\cH_{2\gamma,\text{(had)}}^{(2)}}
        ~.
\end{split}
\end{equation}

\section{Ancillary files}\label{sec:ancillary}

The main new result of the present work is the expression for the hard amplitude $\cH_{2\gamma}^{(2)}$. Let us describe the content of ancillary files:
\begin{enumerate}
\item Hard invariant amplitudes $H_{k,n\gamma}^{(L)}$ with $n=1,2$ and $L=0,1,2$ at $\epsilon = 0$ that are sufficient for obtaining QED contributions to C-odd radiative corrections and asymmetries. Note that all amplitudes except $H_{k,2\gamma}^{(2)}$ are just the same we used in \cite{GKL2025}. The results for the amplitudes together with their various asymptotics are located in \texttt{H/} folder.
\item Integration kernels $H_{k,\Lambda^2}^{(1)}$ that are necessary to evaluate hadronic vacuum polarisation contribution $\mathcal{H}_{2\gamma,\text{(had)}}^{(2)}$ using Eqs. \eqref{eq:HhadIntegral} and \eqref{eq:HLambda2}. The results for the kernels are located in \texttt{HVP/} folder.
\end{enumerate}
Each folder contains the \texttt{Readme.md} file describing the folder content.

\section{Numerical results}
The full discussion should also include polarization effects, which can be readily obtained from the invariant amplitudes computed in this work. However, we restrict ourselves here to the unpolarized case. The Mathematica notebook \texttt{Asymmetry.nb}, which constructs relative corrections from the hard amplitudes and combinations of soft factors $\cV$ and $\cW$, is attached to the paper.

\begin{figure}
	\centering
	\includegraphics[width=7.5cm]{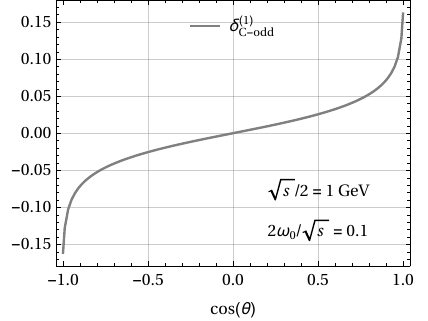}
	\includegraphics[width=7.5cm]{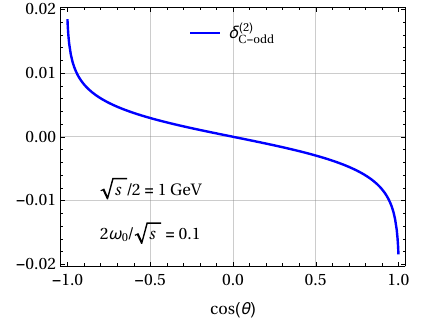}
	\caption{One- and two-loop C-odd radiative corrections, $\delta^{(1)}_{\text{C-odd}}$ and $\delta^{(2)}_{\text{C-odd}}$ for the process $e^{+}e^{-}\to \mu^{+}\mu^{-}$ as functions of $\cos\theta$ for $E_\text{beam}=1$ GeV and $\omega_0=0.1 E_\text{beam}$.}
	\label{fig:asym12}
\end{figure}

The $C$-odd part of radiative corrections $\delta^{(1)}_{C\text{-odd}}$  and  $\delta^{(2)}_{C\text{-odd}}$ are polynomial in two variables, $L_\omega = \ln \frac{\sqrt{s}}{2\omega_0}$ and $L=\ln \frac{s}{m^2}$:
\begin{align}
	\delta^{(1)}_{C\text{-odd}} &= \delta^{(1)}_{10} L_{\omega} + \delta^{(1)}_{00}\,,\quad \delta_{10}^{(1)} = 16a\ln\left(\frac{1+\beta c}{1-\beta c}\right)~,\\
	\delta^{(2)}_{C\text{-odd}} &= \delta_{21}^{(2)} L_\omega^2 L + \sum_{k,n = 0}^{k + n \le 2}\delta^{(2)}_{kn}L_{\omega}^{k}L^{n},\quad \delta_{21}^{(2)} = -128 a^2 \ln \left(\frac{1+\beta c}{1-\beta c}\right)
	\label{eq:delta2_form}
    ~,
\end{align}
with coefficients $\delta_{kn}^{(1)}$ and $\delta_{kn}^{(2)}$ being the functions of $\beta=\sqrt{1-4M^2/s}$ and $c=\cos\theta$. The coefficients are asymmetric under $c\to -c$, reflecting the angular asymmetry. The computed one-loop asymmetry $\delta^{(1)}_{C\text{-odd}}$ numerically coincides with the \cite{Berends1973}. For the  beam energy $E_\text{beam}=1\,\text{GeV}$ and $\omega_{0} = 0.1 E_{\text{beam}}$ we have $\delta^{(1)}_{C\text{-odd}}$ is of the order of $10\%$, while two-loop asymmetry $\delta^{(2)}_{C\text{-odd}}$ is of the order of $1\%$, see Fig.~\ref{fig:asym12}. The two-loop correction has the opposite sign and approximately the same shape as one-loop contribution.

\begin{figure}
	\centering
	\includegraphics[width=7.5cm]{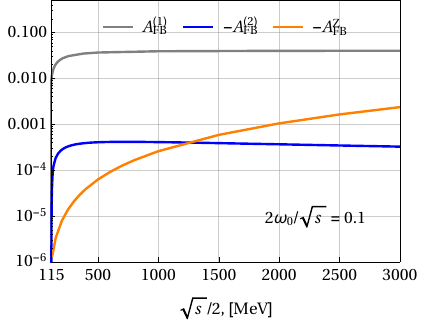}
	\put(-110,32){\makebox(0,0)[lb]{\bf(a)}}
	\includegraphics[width=7.5cm]{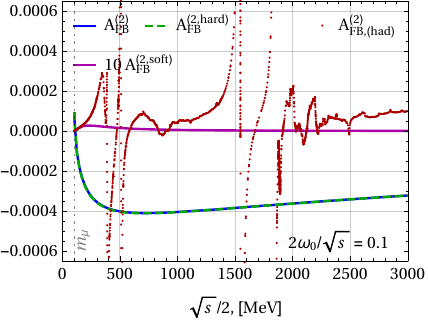}
	\put(-110,32){\makebox(0,0)[lb]{\bf(b)}}
	\caption{(a) $A_{\text{FB}}^{(1)}$ and $A_{\text{FB}}^{(2)}$ --- QED contributions to forward-backward asymmetry in the process $e^{+}e^{-} \to \mu^{+}\mu^{-}$ as functions of electron beam energy $E_\text{beam} = \sqrt{s}/2$ for $\omega_{0}=0.1 E_\text{beam}$; $A_{\text{FB}}^{\text{Z}}$ --- $Z$-boson exchange contribution at tree level. (b) Two-loop forward-backward asymmetry: QED~contributions $A_{\text{FB}}^{(2)} = A_{\text{FB}}^{(2,\text{hard})} + A_\text{FB}^{(2,\text{soft})}$ and hadronic vacuum polarization contribution $A_{\text{FB,(had)}}^{(2)}$.}
	\label{fig:asymfb}
\end{figure}

One- and two-loop contributions to forward-backward asymmetry are shown on the Fig.~\ref{fig:asymfb}(a) along with an estimate of $Z$-boson contribution, which is evaluated at the tree level and assuming that the total energy is far from the $Z$-boson mass, $s \ll M_Z^2$:
\begin{equation}
  A_{\text{FB}}^{Z} \approx 
  -\frac{3 \beta s}{16(3-\beta^2)\cos^2{\theta_W}\sin^2{\theta_W}(M_Z^2-s)}
  ~,
\end{equation}
where for the numerical values we use the $Z$-boson mass $M_Z=91.2\ \text{GeV}/c^2$, and the sine of Weinberg angle $\sin{\theta_W}=0.223$. The two-loop contribution to forward-backward asymmetry, $A_{\text{FB}}^{(2)}$, appeared at the level of $10^{-4}$. 
The two contributions $A_\text{FB}^{(2,\text{soft})}$ and $A_\text{FB}^{(2,\text{hard})}$ to forward-backward asymmetry at the two-loop level are shown on the Fig.~\ref{fig:asymfb}(b). Only~$A_\text{FB}^{(2,\text{soft})}$ depends on $\omega_{0}$, and moreover, its contribution is small in the whole energy range. The finite value $A_{\text{FB}}^{(2)}$ at the muon pair production threshold turned out to be equal $A_{\text{FB}}^{(2)}(s=4m_{\mu}^2) = \frac{5}{3}\alpha^2$. The hadronic vacuum polarization contribution $A_{\text{FB,(had)}}^{(2)}$ is also shown on the Fig.~\ref{fig:asymfb}(b), it reflects peculiar properties of experimental data on hadronic cross sections\footnote{	For numerical estimates we used the parametrization of the hadronic polarization operator described in \cite{Ignatov2008} and the source code provided at \url{https://cmd.inp.nsk.su/~ignatov/vpl/}} and its behavior near resonances deserves a special dedicated consideration.

\section{Conclusion}
\label{sec:conclusion}
To conclude, we have calculated all QED and NNLO contributions to the $C$-odd part of the differential cross section, including the hadronic vacuum polarization insertion,  for \eemumu process. This part of the cross section corresponds to the angular asymmetry in comparison to $\theta\to\pi-\theta$. We used the fact that the ratio $m^2/M^2=m_e^2/m_\mu^2$ is tiny and neglected the power corrections in this parameter. Our results can be used for arbitrary polarization of all involved particles.


\appendix

\section{Master integrals for muonic contribution to $2\gamma$-reducible amplitude}
\label{sec:basis-Lmu-1}

\begin{figure}
  \centering
  \includegraphics[width=.8\textwidth]{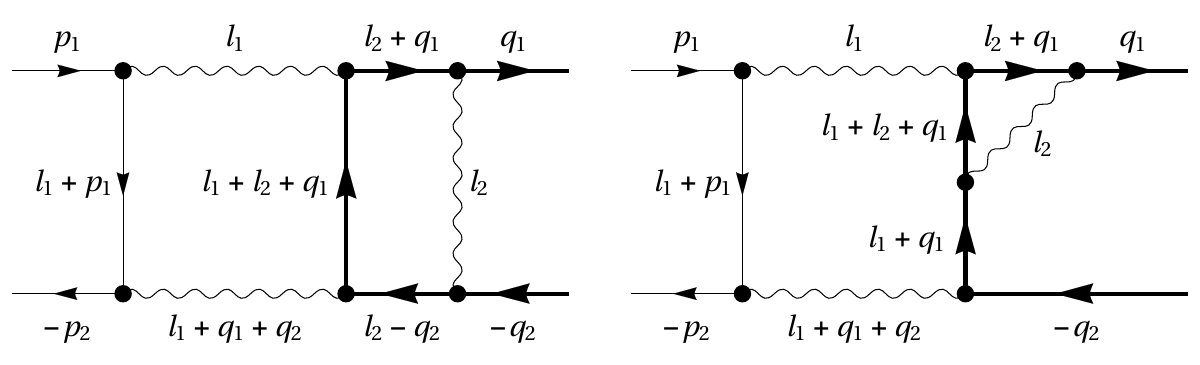}
  \caption{Two Feynman diagrams that generate families of integrals; $l_{1}, l_{2}$ --- loop momenta, charged particles lines are labeled with the momentum flow along the arrow.}
  \label{fig:d2}
\end{figure}
Feynman diagrams of the two different topologies are presented in Fig.~\ref{fig:d2}. The corresponding integral families can be written as
\begin{equation}
  \label{eq:integral-families}
  j(n_{1}, \dots, n_{9}) =
   \int \frac{d^{d}l_{1}}{i \pi^{d/2}}\, \frac{d^{d}l_{2}}{i \pi^{d/2}} \
  \prod_{k=1}^{9} \frac{1}{D_{k}^{n_{k}}}~,
\end{equation}
where
\begin{equation}
  \begin{split}
  \label{eq:denominators}
    &D_{1}=-l_{2}^2~,\quad
    D_{2}=M^2-\left(l_{2}+q_{1}\right){}^2~,\quad
    D_{3}=M^2-\left(l_{1}+l_{2}+q_{1}\right){}^2~,\quad\\
    &D_{4}=M^2-\left(l_{2}-q_{2}\right){}^2~,\quad
    D_{5}=-l_{1}^2~,\quad
    D_{6}=-\left(l_{1}+p_{1}\right){}^2~,\quad\\
    &D_{7}=-\left(l_{1}+q_{1}+q_{2}\right){}^2~,\quad
    D_{8}=M^2-\left(l_{1}+q_{1}\right){}^2~,\quad
    D_{9}=l_{2}\cdot p_{1}~.
  \end{split}
\end{equation}
The last combination, $D_{9}$, is added as irreducible numerator, so that $n_9 \leq 0$. Only one of the indices $n_{4}$ or $n_{8}$ can be positive depending on the generating topology. Electron mass is set to zero from the beginning, $m = 0$.

Using \LiteRed~\cite{Lee:2013mka} we perform IBP reduction, and get the basis of $42$ master integrals, $\vec{j}=\{j_{1},\dots,j_{42}\}^{T}$. Among them $32$ master integrals come from the first family and $10$ are additional master integrals for the second family. The differential equations have the form:
\begin{equation}
  \partial_{s} \vec{j} = M_{s} \vec{j}~,\quad
  \partial_{t} \vec{j} = M_{t} \vec{j}~,
\end{equation}
where $M_{s}$, $M_{t}$ are rational matrices depending on $s$, $t$ and $\epsilon$.

We use \Libra~\cite{Lee:2020zfb} for transformations of the differential systems. In order to get $\epsilon$-form we introduce variables $\beta$ and $c = \cos(\theta)$
and get the transformation matrix $\vec{j} = T \vec{J}$ that reduce both systems to $\epsilon$-form:
\begin{equation}
  \label{eq:eps-form}
  \partial_{\beta} \vec{J} = \epsilon M_{\beta} \vec{J}~,\quad
  \partial_{c} \vec{J} = \epsilon M_{c} \vec{J}~.
\end{equation}

Any specific solution has the form $\vec{J} = U \vec{J}_{0}$, where $U$ is an evolution matrix and $J_{0}$ are boundary constants. The matrix $U$ is defined by the concrete evolution path from the boundary. Using matrices $M_{\beta}$ and $M_{c}$ in \eqref{eq:eps-form}, \Libra constructs an expansion of the matrix $U$ with respect to $\epsilon$ up to the desired order and finds relations between $\vec{J}_{0}$ and specific asymptotic coefficients $\vec{c}_{0}$ of the original integrals $\vec{j}$, $\vec{J}_{0} = L \vec{c}_{0}$, so that we have
\begin{equation}
  \vec{J}_{0} = U L_{0} \vec{c}_{0}~.
\end{equation}
It is possible to set boundary conditions at different points and perform evolution to the necessary point $\{\beta,~c\}$. The threshold point $\beta = 0$ is convenient, since there is no dependence of the asymptotic coefficients $\vec{c}_{0}$ on $\cos(\theta)$. We managed to evaluate $34$ out of $42$ necessary constants at the limit $\beta = 0$ (i.e. we can express them in terms specific values of hypergeometric functions with indices depending on $\epsilon$). At this point it is possible to find a common factor that makes all $\vec{J}$ uniformly transcendental and include this factor to the transformation $T$.

The remaining $8$ asymptotic constants were found by matching with the solution coming from another boundary. Namely, we consider the limit $\{\beta = 1,~c = 0\}$, evaluate necessary asymptotic constants $\vec{c}_{1}$ for boundary conditions $\vec{J}_{1} = L_{1} \vec{c}_{1}$. Then we  obtain an expansion of the associator $U_{10}$ that perform evolution with respect to $\beta$ from $0$ to $1$ at fixed $c=0$ so that
\begin{equation}
  \label{eq:associator}
  L_{1}\vec{c}_{1} = U_{10}L_{0}\vec{c}_{0}~.
\end{equation}
Some of the new constants $\vec{c}_{1}$ also can not be easily evaluated, but the Eq.~\eqref{eq:associator} yields a number of relations between expansions of the constants at the two different limits, that allow to restore expansions for the remaining 8 constants $\vec{c}_{0}$ at $\beta = 0$.

We have checked non-trivial matching relations from the different limits in the Eq.~\eqref{eq:associator}. We have also performed cross-checks of numerical values of the master integrals obtained using \texttt{FIESTA} \cite{Smirnov:2021rhf}.

The final result for master integrals $\vec{J}$ were obtained in the form of the expansion with respect to $\epsilon$ up to $O(\epsilon^{6})$ with the coefficients $G(\vec{w}|\beta)$ with the weights  $w_{i} \in \{0,\pm 1, \pm \frac{1}{c}, c, e^{\pm i \theta} = c\pm i \sqrt{1-c^2}\}$.

\bibliographystyle{JHEP}
\bibliography{as_eemumu}

@Article{Berends1973,
  author    = {Berends, Frits A and Gaemers, KJF and Gastmans, Raymond},
  journal   = {Nuclear Physics B},
  title     = {$\alpha$3-contribution to the angular asymmetry in e+ e-→ $\mu$+ $\mu$-},
  year      = {1973},
  pages     = {381--397},
  volume    = {63},
  publisher = {Elsevier},
}

@Article{Berends1983,
  author  = {Berends, Frits A and Kleiss, R and Jadach, S and Was, Z},
  journal = {Acta Phys. Pol., Series B;(Poland)},
  title   = {QED radiative corrections to electron-positron annihilation into heavy fermions},
  year    = {1983},
  number  = {6},
  volume  = {14},
}

@Article{Jadach1984,
  author  = {Jadach, Stanislaw and Was, Zbigniew},
  journal = {Acta Physica Polonica. Series B},
  title   = {QED $0(\alpha^3)$ radiative corrections to the reaction e+ e-→ $\tau$+ $\tau$-including spin and mass effects},
  year    = {1984},
  number  = {12},
  pages   = {1151--1184},
  volume  = {15},
}

@Article{Aliberti2024,
  author        = {Aliberti, Riccardo and others},
  title         = {{Radiative corrections and Monte Carlo tools for low-energy hadronic cross sections in $e^+ e^-$ collisions}},
  year          = {2024},
  month         = {10},
  archiveprefix = {arXiv},
  eprint        = {2410.22882},
  primaryclass  = {hep-ph},
}

@Article{Lee2025,
  author        = {Lee, Roman N.},
  journal       = {J. High Energy Phys.},
  title         = {{Two-loop master integrals for $e^+e^- \to \mu^+\mu^-$ process with account of electron mass}},
  year          = {2025},
  pages         = {006},
  volume        = {02},
  archiveprefix = {arXiv},
  doi           = {10.1007/JHEP02(2025)006},
  eprint        = {2412.00793},
  primaryclass  = {hep-ph},
}

@Article{GKL2025,
  author        = {Gerasimov, Roman E. and Krachkov, Petr A. and Lee, Roman N.},
  journal       = {J. High Energy Phys.},
  title         = {{Electron-positron annihilation into heavy leptons at two loops}},
  year          = {2025},
  pages         = {118},
  volume        = {08},
  archiveprefix = {arXiv},
  doi           = {10.1007/JHEP08(2025)118},
  eprint        = {2503.09245},
  primaryclass  = {hep-ph},
}

@article{Lee:2013mka,
    author = "Lee, Roman N.",
    editor = "Wang, Jianxiong",
    title = "{LiteRed 1.4: a powerful tool for reduction of multiloop integrals}",
    eprint = "1310.1145",
    archivePrefix = "arXiv",
    primaryClass = "hep-ph",
    doi = "10.1088/1742-6596/523/1/012059",
    journal = "J. Phys. Conf. Ser.",
    volume = "523",
    pages = "012059",
    year = "2014"
}

@article{Lee:2020zfb,
    author = "Lee, Roman N.",
    title = "{Libra: A package for transformation of differential systems for multiloop integrals}",
    eprint = "2012.00279",
    archivePrefix = "arXiv",
    primaryClass = "hep-ph",
    doi = "10.1016/j.cpc.2021.108058",
    journal = "Comput. Phys. Commun.",
    volume = "267",
    pages = "108058",
    year = "2021"
}

@article{Ignatov:2022iou,
    author = "Ignatov, Fedor and Lee, Roman N.",
    title = "{Charge asymmetry in $e^{+}e^{-}\to\pi^{+}\pi^{-}$ process}",
    eprint = "2204.12235",
    archivePrefix = "arXiv",
    primaryClass = "hep-ph",
    doi = "10.1016/j.physletb.2022.137283",
    journal = "Phys. Lett. B",
    volume = "833",
    pages = "137283",
    year = "2022"
}

@Article{Yennie1961,
  author   = {D.R Yennie and S.C Frautschi and H Suura},
  journal  = {Annals of Physics},
  title    = {The infrared divergence phenomena and high-energy processes},
  year     = {1961},
  issn     = {0003-4916},
  number   = {3},
  pages    = {379-452},
  volume   = {13},
  doi      = {https://doi.org/10.1016/0003-4916(61)90151-8},
  url      = {https://www.sciencedirect.com/science/article/pii/0003491661901518},
}

@Article{Fadin:2023phc,
  author        = {Fadin, V. S. and Lee, R. N.},
  journal       = {JHEP},
  title         = {{Two-loop radiative corrections to $e^+e^-\to\gamma\gamma^*$ cross section}},
  year          = {2023},
  pages         = {148},
  volume        = {11},
  archiveprefix = {arXiv},
  doi           = {10.1007/JHEP11(2023)148},
  eprint        = {2308.09479},
  primaryclass  = {hep-ph},
}

@article{Smirnov:2021rhf,
    author = "Smirnov, A. V. and Shapurov, N. D. and Vysotsky, L. I.",
    title = "{FIESTA5: Numerical high-performance Feynman integral evaluation}",
    eprint = "2110.11660",
    archivePrefix = "arXiv",
    primaryClass = "hep-ph",
    doi = "10.1016/j.cpc.2022.108386",
    journal = "Comput. Phys. Commun.",
    volume = "277",
    pages = "108386",
    year = "2022"
}

@phdthesis{Ignatov2008,
  author = "Ignatov, F.V.",
  title = "Measurement of the pion form-factor at 1.04-1.38 GeV energy range with the CMD-2 detector",
  year = "2008",
  school = {BINP, Novosibirsk},
  url = "https://cmd.inp.nsk.su/~ignatov/vpl/",
  institution = "BINP, Novosibirsk"
}

\end{document}